\documentclass[prl, twocolumn]{revtex4}

\usepackage{amsmath}    
\usepackage{graphicx}   
\usepackage{verbatim}   
\usepackage{color}      
\usepackage{subfigure}  
\usepackage{hyperref}   

\begin{document}
\title{Side face excited microstructured fibers for photonic integrated circuits formations}
\author{Igor V. Guryev}
\email{guryev@dulcinea.ugto.mx}

\author{Igor A. Sukhoivanov}

\author{J.-A. Andrade Lucio}

\author{E. Alvarado-Mendez}
\affiliation{DICIS, Universidad de GTo, Salamanca, Mexico}%

\begin{abstract}
In the paper, we propose the new technology for mass production of the photonic crystal devices and all-optical integrated circuits. We carry out the brief analysis of positive and negative sides paying attention to recent advances in silicon fibers fabrication which is crucial moment in the proposed technology.  
\end{abstract}

\maketitle
\section{Introduction}

Nowadays, one of the most important challenges concern mass production of the photonic crystal (PhC) devices. At the moment, there are several kinds of the PhC devices available for mass production. Among them there are the artificial opals \cite{Darragh, Mayoral} including inverted opals \cite{Wijnhoven} and the PhC fibers \cite{Russel}. However, the artificial opals are strictly periodical 3D PhCs with no possibility of the controllable defect introduction and, therefore, cannot be used for the mass production of the PhC devices. On the other hand, the technologies which provides fabrication of the integrated photonic circuits such as electronic beam lithography, are too slow and expensive so they cannot be used for mass production of the devices. 

The technology of the PhC fibers though allows production of the PhC fibers with extremely high accuracy and with any profile the developers can imagine. Such an advantage is proved by dozens of years of production of the convenient fibers for telecommunications. 

\begin{figure}[b]
\includegraphics[width=9cm]{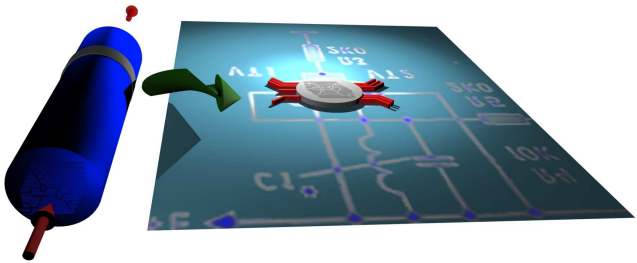}
\caption{Side-excited PCF as a part of the photonic integrated circuit}
\label{fig:PCF_demo}
\end{figure}

Therefore, the main idea of this paper consists in providing the possibility of the mass production by taking advantage of the PCF technology. Usually, the PCF is used by guiding the light which travels along the fiber as shown in the left part of the figure 1. However, it is also possible to use the fiber in a different manner by irradiating it from the side face. In this case, a single fiber with special refractive index profile can be splitted into many pieces. Each piece can be integrated into photonic scheme or can be used as a single photonic scheme (see the right part of the figure \ref{fig:PCF_demo}).

\section{Justification}

\begin{figure}[h]
\subfigure[]{\label{fig:bs_PCF}\includegraphics[width=8cm]{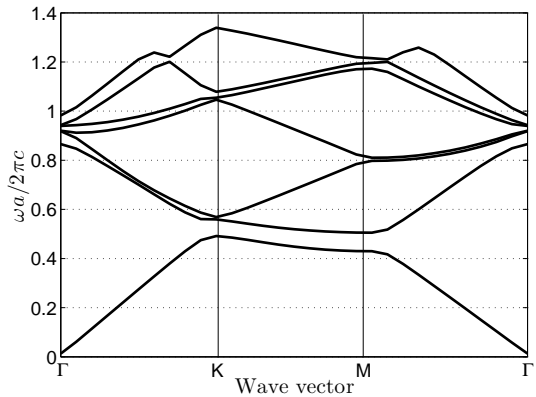}}
\subfigure[]{\label{fig:disp_wg_PCF}\includegraphics[width=8cm]{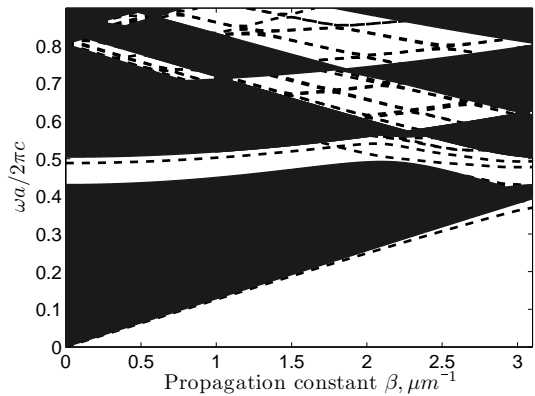}}
\caption{Band structure (a) of the PCF considered as in-plane PhC and the dispersion characteristic (b) of the waveguide in such a PhC}
\label{fig:PhC_characteristics}
\end{figure}

The proposed idea, however, requires several modifications to be done to the PCF fabrication technological process to provide the new way of using the PCF. The most important modification concerns the refractive index of the material the fiber is made of. In case of traditional use, the refractive index contrast in the PCF is enough to provide the light localization inside the defect. However, attempts to observe the light localization when irradiating the fiber from the side face are doomed. Because of the low refractive index contrast, the ptotonic band gap cannot be observed in any kind of 2D photonic crystal which is illustrated in the figure \ref{fig:bs_PCF}. Actually the in-plane band structure of the PhC confining the light within the PCF core is very close to the one of the free space and contains only tiny partial band gaps which are able to provide the effect of light diffraction but not the localization. The PBG was found when computing the orthogonalized band structure of the PhC waveguide created in such a PhC (see the figure \ref{fig:disp_wg_PCF}). However, even though this can provide the light guiding effect, the low refractive index contrast does not allow to create more complex elements for the information processing.

\begin{figure}[h]
\begin{center}
\includegraphics[width=7cm]{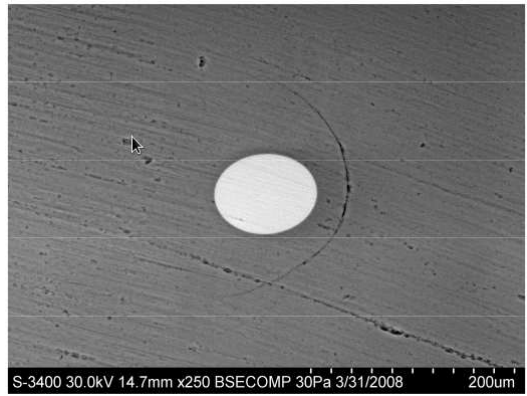}
\end{center}
\caption{The cross-section of the silicon fiber presented in the \cite{Ballato}}
\label{fig:silicon_fiber}
\end{figure}

The properties of such a PhC can be greatly improved by increasing the refractive index. Recently, the technological process of the PCF production was dramatically improved. Moreover, there have been made simple step-index fibers of silicon \cite{Ballato} (see the photo in figure \ref{fig:silicon_fiber}) which refractive index is more than 3 and, therefore, provides high refractive-index contrast. 

Now, the technological process allows production of the step-index silicon fibers with submicron core radius \cite{Healy}. This experience can be used to produce silicon PCFs with the structure of arbitrary complexity and, therefore, provide the possibility for mass-production of the PhC planar devices.

\section{Preliminary analysis of the technology efficiency}

As a new technology, the proposed way of the PhC devices production possesses both advantages and disadvantages. The most important advantages are the following:
\begin{itemize}
\item Providing the ability of mass production. This advantage makes the proposed technology more preferable than well-known etching and lithography techniques. Moreover, it provides high repeatability since within the existing PCF production technology the PhC possesses the same refractive index profile along all the PCF.

\item Providing the new possibilities for the PhC elements' refractive index profile. For instance, with PCF production technique it is possible to create the gradient profile of a single element by applying the gradient fibers technologies at pre-forming stage. Such a complex profile is impossible to create with existing PhC production technology. 
\item The nonlinear materials can be introduced into the PhC elements at pre-forming stage as well as after the fiber formation process is complete. 
\item Guiding properties of the fiber itself can be used to provide the pumping of the active regions of the PhC. This may be accomplished by introducing the light along the PhC elements as it is made in regular fibers.
\item Finally, the PCF technology may provide the production of the all-optical PhC-based circuits of an arbitrary complexity. With this, the whole scheme may be created within the profile of a single PCF. On the other hand, the scheme can be composed of several modules of lower complexity and connected by planar optical waveguides in a single substrate.
\end{itemize}

However, together with all the advantages, there are several limitation of the new technology such as:

\begin{itemize}
\item The silicon fiber production involves new sides of the technological process which, at this moment, limits the quality of the silicon fibers. The most negative effects involve interfaces irregularities which leads to high losses.
\item If the scheme consists of several modules produced by the PCF technology, they need to be connected by planar optical waveguides. However, as is shown in figure 1, the PhC scheme is surrounded by the circular coating. This coating causes big problems when trying to agree the scheme with planar waveguides. The problem can be solved by polishing the faces or by introducing the micro-lenses for the radiation focusing at the waveguides input. However, this still involves additional precise step and increases the complexity of the whole technology.
\item The other problem involving the additional step is vertical confinement. After the fiber is sliced into similar pieces, the only factor providing the vertical confinement is the total internal reflection from the cleaves. However, this situation is unacceptable since vertical dimensions of the fiber slices are much larger than the operating wavelength which will cause additional losses. Possible solutions of this problem is to use fiber tapering, UV fiber deposition or selective oxygenation. By solving the problem, this also introduces additional steps into the process, thus, increasing the price.
\end{itemize}

\section{Conclusion}
In the work, we proposed the new way for the mass production of the PhC devices and integrated scheme. We have briefly described the most important positive and negative sides of the technology as well as have given some ways for the problems solutions which may occur. 

\bibliographystyle{aipnum4-1}
\bibliography{lit}

\begin{thebibliography}{1}%
\makeatletter
\providecommand \@ifxundefined [1]{%
 \ifx #1\undefined \expandafter \@firstoftwo
 \else \expandafter \@secondoftwo
\fi
}%
\providecommand \@ifnum [1]{%
 \ifnum #1\expandafter \@firstoftwo
 \else \expandafter \@secondoftwo
\fi
}%
\providecommand \enquote [1]{``#1''}%
\providecommand \bibnamefont  [1]{#1}%
\providecommand \bibfnamefont [1]{#1}%
\providecommand \citenamefont [1]{#1}%
\providecommand\href[0]{\@sanitize\@href}%
\providecommand\@href[1]{\endgroup\@@startlink{#1}\endgroup\@@href}%
\providecommand\@@href[1]{#1\@@endlink}%
\providecommand \@sanitize [0]{\begingroup\catcode`\&12\catcode`\#12\relax}%
\@ifxundefined \pdfoutput {\@firstoftwo}{%
 \@ifnum{\z@=\pdfoutput}{\@firstoftwo}{\@secondoftwo}%
}{%
 \providecommand\@@startlink[1]{\leavevmode\special{html:<a href="#1">}}%
 \providecommand\@@endlink[0]{\special{html:</a>}}%
}{%
 \providecommand\@@startlink[1]{%
  \leavevmode
  \pdfstartlink
   attr{/Border[0 0 1 ]/H/I/C[0 1 1]}%
   user{/Subtype/Link/A<</Type/Action/S/URI/URI(#1)>>}%
  \relax
 }%
 \providecommand\@@endlink[0]{\pdfendlink}%
}%
\providecommand \url  [0]{\begingroup\@sanitize \@url }%
\providecommand \@url [1]{\endgroup\@href {#1}{\urlprefix}}%
\providecommand \urlprefix [0]{URL }%
\providecommand \Eprint[0]{\href }%
\@ifxundefined \urlstyle {%
  \providecommand \doi [1]{doi:\discretionary{}{}{}#1}%
}{%
  \providecommand \doi [0]{doi:\discretionary{}{}{}\begingroup
  \urlstyle{rm}\Url }%
}%
\providecommand \doibase [0]{http://dx.doi.org/}%
\providecommand \Doi[1]{\href{\doibase#1}}%
\providecommand \selectlanguage [0]{\@gobble}%
\providecommand \bibinfo [0]{\@secondoftwo}%
\providecommand \bibfield [0]{\@secondoftwo}%
\providecommand \translation [1]{[#1]}%
\providecommand \BibitemOpen[0]{}%
\providecommand \bibitemStop [0]{}%
\providecommand \bibitemNoStop [0]{.\EOS\space}%
\providecommand \EOS [0]{\spacefactor3000\relax}%
\providecommand \BibitemShut [1]{\csname bibitem#1\endcsname}%
\bibitem{Darragh}%
  \BibitemOpen
  \bibfield{author}{%
  \bibinfo {author} {\bibfnamefont{P.~J.}\ \bibnamefont{Darragh}}\ and\
  \bibinfo {author} {\bibfnamefont{J.~L.}\ \bibnamefont{Perdrix}},\ }%
  \bibfield{journal}{%
  \bibinfo {journal} {J. Gemmol}\ }%
  \textbf{\bibinfo {volume} {14}},\ \bibinfo {pages} {215} (\bibinfo {year}
  {1975})\BibitemShut{NoStop}%
\bibitem{Mayoral}%
  \BibitemOpen
  \bibfield{author}{%
  \bibinfo {author} {\bibfnamefont{R.}~\bibnamefont{Mayoral}}, \bibinfo
  {author} {\bibfnamefont{J.}~\bibnamefont{Requena}}, \bibinfo {author}
  {\bibfnamefont{C.}~\bibnamefont{L{\'o}pez}}, \bibinfo {author}
  {\bibfnamefont{S.}~\bibnamefont{Moya}}, \bibinfo {author}
  {\bibfnamefont{H.}~\bibnamefont{M{\'i}guez}}, \bibinfo {author}
  {\bibfnamefont{L.}~\bibnamefont{V{\'a}zquez}}, \bibinfo {author}
  {\bibfnamefont{F.}~\bibnamefont{Meseguer}}, \bibinfo {author}
  {\bibfnamefont{M.}~\bibnamefont{Holgado}}, \bibinfo {author}
  {\bibfnamefont{A.}~\bibnamefont{Cintas}},\ and\ \bibinfo {author}
  {\bibfnamefont{A.}~\bibnamefont{Blanco}},\ }%
  \bibfield{journal}{%
  \bibinfo {journal} {Adv. Mater.}\ }%
  \textbf{\bibinfo {volume} {9}},\ \bibinfo {pages} {257} (\bibinfo {year}
  {1997})\BibitemShut{NoStop}%
\bibitem{Wijnhoven}%
  \BibitemOpen
  \bibfield{author}{%
  \bibinfo {author} {\bibfnamefont{J.}~\bibnamefont{Wijnhoven}}, \bibinfo
  {author} {\bibfnamefont{L.}~\bibnamefont{Bechger}},\ and\ \bibinfo {author}
  {\bibfnamefont{W.}~\bibnamefont{Vos}},\ }%
  \bibfield{journal}{%
  \bibinfo {journal} {Chem. Mater.}\ }%
  \textbf{\bibinfo {volume} {13}},\ \bibinfo {pages} {4486} (\bibinfo {year}
  {2001})\BibitemShut{NoStop}%
\bibitem{Russel}%
  \BibitemOpen
  \bibfield{author}{%
  \bibinfo {author} {\bibfnamefont{P.~S.~J.}\ \bibnamefont{Russell}},\ }%
  \bibfield{journal}{%
  \bibinfo {journal} {Science}\ }%
  \textbf{\bibinfo {volume} {299}},\ \bibinfo {pages} {358} (\bibinfo {year}
  {2003})\BibitemShut{NoStop}%
\bibitem{Ballato}%
  \BibitemOpen
  \bibfield{author}{%
  \bibinfo {author} {\bibfnamefont{J.}~\bibnamefont{Ballato}}, \bibinfo
  {author} {\bibfnamefont{T.}~\bibnamefont{Hawkins}},\ and\ \bibinfo {author}
  {\bibfnamefont{P.}~\bibnamefont{Foy}},\ }%
  \bibfield{journal}{%
  \bibinfo {journal} {Optics Express}\ }%
  \textbf{\bibinfo {volume} {16}},\ \bibinfo {pages} {18675} (\bibinfo {year}
  {2008})\BibitemShut{NoStop}%
\bibitem{Healy}%
  \BibitemOpen
  \bibfield{author}{%
  \bibinfo {author} {\bibfnamefont{N.}~\bibnamefont{Healy}}, \bibinfo {author}
  {\bibfnamefont{J.~R.~S.}\ \bibnamefont{P.}},\ and\ \bibinfo {author}
  {\bibfnamefont{J.~A.}\ \bibnamefont{Sazio}},\ }%
  \bibfield{journal}{%
  \bibinfo {journal} {Optics Express}\ }%
  \textbf{\bibinfo {volume} {18}},\ \bibinfo {pages} {7596} (\bibinfo {year}
  {2010})\BibitemShut{NoStop}%
\end{thebibliography}%
\end{document}